\documentclass[11pt]{article}

\usepackage{fullpage,latexsym,amsmath,amsfonts}

\def\01{\{0,1\}}
\newcommand{\ceil}[1]{\lceil{#1}\rceil} 

\newcommand{\eps}{\varepsilon}

\newcommand{\Exp}{\mbox{\rm Exp}}
\newcommand{\LDC}{\mbox{\rm LDC}}
\newcommand{\MEM}{\mbox{\rm MEM}}
\newcommand{\IP}{\mbox{\rm IP}}
\newcommand{\norm}[1]{\mbox{$\parallel{#1}\parallel$}}
\newtheorem{definition}{Definition}
\newtheorem{theorem}{Theorem}
\newtheorem{lemma}{Lemma}

\newenvironment{proof}
{\noindent {\bf Proof. }}
{{\hfill $\Box$}\\
 \smallskip}

\begin{document}

\title{Error-Correcting Data Structures}
\author{
Ronald de Wolf\thanks{rdewolf@cwi.nl. Partially supported by Veni and Vidi grants from the Netherlands Organization for Scientific Research (NWO), and by the European Commission under the Integrated Project Qubit Applications (QAP) funded by the IST directorate
as Contract Number 015848.}\\
CWI Amsterdam}
\date{}
\maketitle

\begin{abstract}
We study data structures in the presence of adversarial noise.
We want to encode a given object in a succinct data structure that enables us to efficiently 
answer specific queries about the object, even if the data structure has been corrupted by a constant
fraction of errors.  This new model is the common generalization of (static) data structures and locally 
decodable error-correcting codes. The main issue is the tradeoff between the space used by the data 
structure and the time (number of probes) needed to answer a query about the encoded object. 
We prove a number of upper and lower bounds on various natural error-correcting data structure problems.
In particular, we show that the optimal length of error-correcting data structures
for the {\sc Membership} problem (where we want to store subsets of size $s$ from a universe
of size $n$) is closely related to the optimal length of locally decodable codes for $s$-bit strings.\\[2mm]
\end{abstract}


\section{Introduction}


Data structures deal with one of the most fundamental questions of computer science:
how can we store certain objects in a way that is both space-efficient and that enables
us to efficiently answer questions about the object?
Thus, for instance, it makes sense to store a set as an ordered list or as a heap-structure, 
because this is space-efficient and allows us to determine quickly (in time logarithmic in the size of the set)
whether a certain element is in the set or not.

From a complexity-theoretic point of view, the aim is usually to study the tradeoff between
the two main resources of the data structure: the length/size of the data structure (storage space) 
and the efficiency with which we can answer specific queries about the stored object.
To make this precise, we measure the length of the data structure in bits, and measure the 
efficiency of query-answering in the number of \emph{probes}, i.e., the number of bit-positions
in the data structure that we look at in order to answer a query.
The following is adapted from Miltersen's survey~\cite{miltersen:cellprobesurvey}:

\begin{definition}\label{defdatastr}
Let $D$ be a set of data items, $Q$ be a set of queries, $A$ be a set of answers, and
$f:D\times Q\rightarrow A$.
A \emph{$(p,\eps)$-data structure} for $f$ of length $N$ is a map $\phi:D\rightarrow\01^N$ for which there is 
a randomized algorithm $\cal A$ that makes at most $p$ probes to its oracle and 
satisfies
for every $q\in Q$ and $x\in D$
$$
\Pr[{\cal A}^{\phi(x)}(q)=f(x,q)]\geq 1-\eps.
$$
\end{definition}

Usually we will study the case $D\subseteq\01^n$ and $A=\01$. Most standard data structures 
taught in undergraduate computer science are deterministic, and hence have error probability $\eps=0$.
As mentioned, the main complexity issue here is the tradeoff between $N$ and $p$.
Some data structure problems that we will consider are the following:
\begin{itemize}
\item {\sc Equality.} 
$D=Q=\01^n$, and $f(x,y)=1$ if $x=y$, $f(x,y)=0$ if $x\neq y$. This is not a terribly interesting data structure 
problem in itself, since for every $x$ there is only one query $y$ for which the answer is `1'; we merely mention this data structure here because it will be used to illustrate some definitions later on.  
\item {\sc Membership.}
$D=\{x \in\01^n : \mbox{Hamming weight }|x|\leq s\}$, $Q=[n]:=\{1,\ldots,n\}$, and $f(x,i)=x_i$.  In other words, $x$ corresponds to a set 
of size at most $s$ from a universe of size $n$, and we want to store the set in a way that easily 
allows us to make membership queries. This is probably the most basic and widely-studied data 
structure problem of them all~\cite{fks:sparsetable,yao:tables,bmrv:bitvectorsj,rsv:qsetmembershipj}.
Note that for $s=1$ this is {\sc Equality} on $\log n$ bits, while for $s=n$ it is the general
{\sc Membership} problem without constraints on the set.
\item {\sc Substring.}
$D=\01^n$, $Q=\{y\in\01^n : |y|\leq r\}$, $f(x,y)=x_y$, where $x_y$ is the $|y|$-bit substring of $x$ 
indexed by the 1-bits of $y$ (e.g., $1010_{0110}=01$). For $r=1$ it is {\sc Membership}.
\item {\sc Inner product ($\IP_{n,r}$).}
$D=\01^n$, $Q=\{y\in\01^n : |y|\leq r\}$ and $f(x,y)=x\cdot y$~mod~2.
This problem is among the hardest Boolean problems where the answer depends 
on at most~$r$ bits of $x$ (again, for $r=1$ it is {\sc Membership}).
\end{itemize}
More complicated data structure problems such as {\sc Rank}, {\sc Predecessor}, {\sc Nearest neighbor}
have also been studied a lot, but we will not consider them here. 

One issue that the above definition ignores, is the issue of \emph{noise}.
Memory and storage devices are not perfect: the world is full of cosmic rays, small earthquakes, 
random (quantum) events, bypassing trams, etc., that can cause a few errors here and there.
Another potential source of noise is transmission of the data structure over some noisy channel.
Of course, better hardware can partly mitigate these effects, but in many situations it is
realistic to expect a small fraction of the bits in the storage space to become corrupted over time.
Our goal in this paper is to study \emph{error-correcting} data structures.
These still enable efficient computation of $f(x,q)$ from the stored data structure $\phi(x)$, 
even if the latter has been corrupted by a constant fraction of errors.
In analogy with the usual setting for error-correcting codes~\cite{MacWilliamsS77,lint:ecc}, 
we will take a pessimistic, adversarial view of errors here:  
we want to be able to deal with a constant fraction of errors \emph{no matter where they are placed}.
Formally, we define error-correcting data structures as follows.

\begin{definition}\label{defftdatastr}
Let $D$ be a set of data items, $Q$ be a set of queries, $A$ be a set of answers, and
$f:D\times Q\rightarrow A$.
A \emph{$(p,\delta,\eps)$-error-correcting data structure} for $f$ of length $N$ is a map $\phi:D\rightarrow\01^N$ 
for which there is a randomized algorithm $\cal A$ that makes at most $p$ probes to its oracle and satisfies
$$
\Pr[{\cal A}^{y}(q)=f(x,q)]\geq 1-\eps,
$$
for every $q\in Q$, every $x\in D$, and every $y\in\01^N$ at Hamming distance $\Delta(y,\phi(x))\leq\delta N$.
\end{definition}

Definition~\ref{defdatastr} is the special case of Definition~\ref{defftdatastr} where $\delta=0$.%
\footnote{As \cite[end of Section~1.1]{bmrv:bitvectorsj} notes, a data structure can be viewed as locally decodable source
code. With this information-theoretic point of view, an \emph{error-correcting} data structure is a locally decodable
combined source-\emph{channel} code, and our results for {\sc Membership} show that one can sometimes do better than
combining the best source code with the best channel code. We thank one of the anonymous referees for pointing this out.}
Note that if $\delta>0$ then the adversary can always set the
errors in a way that gives the decoder $\cal A$ a non-zero error probability. Hence
the setting with bounded error probability is the natural one for error-correcting data structures.
This contrasts with the standard noiseless setting, where one usually considers deterministic structures.

A simple example of an efficient error-correcting data structure is for {\sc Equality}:
encode $x$ with a good error-correcting code $\phi(x)$.  Then $N=O(n)$, and we can decode by one
probe: given $y$, probe $\phi(x)_j$ for uniformly chosen $j\in[N]$, compare it with $\phi(y)_j$, 
and output 1 iff these two bits are equal. If up to a $\delta$-fraction of the bits in $\phi(x)$ are corrupted,
then we will give the correct answer with probability $1-\delta$ in the case $x=y$. 
If the distance between any two codewords is close to $N/2$ (which is true for instance
for a random linear code), then we will give the correct answer with probability 
about $1/2-\delta$ in the case $x\neq y$.
These two probabilities can be balanced to 2-sided error $\eps=1/3+2\delta/3$.
The error can be reduced further by allowing more than one probe.

We only deal with so-called \emph{static} data structures here: 
we do not worry about updating the $x$ that we are encoding.
What about \emph{dynamic} data structures, which allow efficient updates as well 
as efficient queries to the encoded object? 
Note that if data-items $x$ and $x'$ are distinguishable in the sense that $f(x,q)\neq f(x',q)$ for at least one query $q\in Q$, 
then their respective error-correcting encodings $\phi(x)$ and $\phi(x')$ will have distance $\Omega(N)$.%
\footnote{Hence if all pairs $x,x'\in D$ are distinguishable (which is usually the case),
then $\phi$ is an error-correcting code.}
Hence updating the encoded data from $x$ to $x'$ will require $\Omega(N)$ changes in the data structure,
which shows that a dynamical version of our model of error-correcting data structures 
with efficient updates is not possible.

Error-correcting data structures not only generalize the standard (static) data structures 
(Definition~\ref{defdatastr}), but they also generalize \emph{locally decodable codes}.  
These are defined as follows:

\begin{definition}\label{defldc}
A \emph{$(p,\delta,\eps)$-locally decodable code} (LDC) of length $N$ is a map $\phi:\01^n\rightarrow\01^N$ 
for which there is a randomized algorithm $\cal A$ that makes at most $p$ probes to its oracle and satisfies
$$
\Pr[{\cal A}^{y}(i)=x_i]\geq 1-\eps,
$$
for every $i\in[n]$, every $x\in\01^n$, and every $y\in\01^N$ at Hamming distance $\Delta(y,\phi(x))\leq\delta N$.
\end{definition}

Note that a $(p,\delta,\eps)$-error-correcting data structure for {\sc Membership} (with $s=n$) 
is exactly a $(p,\delta,\eps)$-locally decodable code. 
Much work has been done on LDCs, but their length-vs-probes tradeoff is still
largely unknown for $p\geq 3$. We refer to~\cite{trevisan:eccsurvey} and the references therein.

LDCs address only a very simple type of data structure problem: we have an $n$-bit ``database'' 
and want to be able to retrieve individual bits from it. In practice, databases have more structure and complexity,
and one usually asks more complicated queries, such as retrieving all records within a certain range.
Our more general notion of error-correcting data structures enables a study of such more practical
data structure problems in the presence of adversarial noise.

\paragraph{Comment on terminology.}
The terminologies used in the data-structure and LDC-literature conflict at various points,
and we needed to reconcile them somehow.  
To avoid confusion, let us repeat here the choices we made. 
We reserve the term ``query'' for the 
question $q$ one asks about the encoded data $x$, while accesses to bits of the data structure are called ``probes''
(in contrast, these are usually called ``queries'' in the LDC-literature). 
The number of probes is denoted by $p$.
We use $n$ for the number of bits of the data item~$x$ (in contrast with the literature about {\sc Membership}, 
which mostly uses $m$ for the size of the universe and $n$ for the size of the set). 
We use $N$ for the length of the data structure (while the LDC-literature mostly uses $m$,
except for Yekhanin~\cite{yekhanin:3ldc} who uses $N$ as we do).
We use the term ``decoder'' for the algorithm $\cal A$.
Another issue is that $\eps$ is sometimes used as the error probability 
(in which case one wants $\eps\approx 0$), and sometimes as the bias away 
from 1/2 (in which case one wants $\eps\approx 1/2$).  We use the former.

\subsection{Our results}

If one subscribes to the approach towards errors taken in the area of error-correcting 
codes, then our definition of error-correcting data 
structures seems a very natural one.  Yet, to our knowledge, this definition is new and
has not been studied before (see Section~\ref{ssecrelated} for other approaches). 

\subsubsection{{\sc Membership}}
The most basic data structure problem is probably the {\sc Membership} problem.
Fortunately, our main positive result for error-correcting data structures applies to this problem.

Fix some number of probes $p$, noise level $\delta$, and allowed error probability $\eps$, 
and consider the minimal length of $p$-probe error-correcting data structures for $s$-out-of-$n$ {\sc Membership}. 
Let us call this minimal length $\MEM(p,s,n)$. 
A first observation is that such a data structure is actually a locally decodable code for $s$ bits:
just restrict attention to $n$-bit strings whose last $n-s$ bits are all 0. Hence, with $\LDC(p,s)$ denoting
the minimal length among all $p$-probe LDCs that encode $s$ bits (for our fixed $\eps,\delta$),
we immediately get the obvious lower bound
$$
\LDC(p,s)\leq \MEM(p,s,n).
$$
This bound is close to optimal if $s\approx n$.
Another trivial lower bound comes from the observation that our data structure for {\sc Membership}
is a map with domain of size $B(n,s):=\sum_{i=0}^s{n\choose i}$ and range of size $2^N$ that has to be injective.
Hence we get another obvious lower bound
$$
\Omega(s\log(n/s))\leq \log B(n,s)\leq \MEM(p,s,n).
$$
What about \emph{upper} bounds?
Something that one can always do to construct error-correcting data structures for any problem,
is to take the optimal non-error-correcting $p_1$-probe construction and encode it with 
a $p_2$-probe LDC. If the error probability of the LDC is much smaller than $1/p_1$, then we
can just run the decoder for the non-error-correcting structure, replacing each of its $p_1$ probes by $p_2$ probes
to the LDC.  This gives an error-correcting data structure with $p=p_1p_2$ probes.
In the case of {\sc Membership}, the optimal non-error-correcting data structure of 
Buhrman et al.~\cite{bmrv:bitvectorsj} uses only 1 probe and $O(s\log n)$ bits.
Encoding this with the best possible $p$-probe LDC gives error-correcting data structures
for {\sc Membership} of length $\LDC(p,O(s\log n))$.
For instance for $p=2$ we can use the Hadamard code\footnote{The Hadamard code of $x\in\01^s$ 
is the codeword of length $2^s$ obtained by concatenating the bits $x\cdot y$ (mod 2) for all $y\in\01^s$.
It can be decoded by two probes, since for every $y\in\01^s$ we have $(x\cdot y)\oplus(x\cdot (y\oplus e_i))=x_i$.
Picking $y$ at random, decoding from a $\delta$-corrupted codeword will be correct with probability at least $1-2\delta$,
because both probes $y$ and $y\oplus e_i$ are individually random and hence probe a corrupted entry 
with probability at most $\delta$. This exponential length is optimal for 2-probe LDCs~\cite{kerenidis&wolf:qldcj}.}
for $s$ bits, giving upper bound $\MEM(2,s,n)\leq \exp(O(s\log n))$.

Our main positive result in Section~\ref{secmembership} says that something much better 
is possible---the max of the above two lower bounds is not far from optimal. 
Slightly simplifying\footnote{Our actual result, Theorem~\ref{thldctoftmem}, 
is a bit dirtier, with some deterioration in the error and noise parameters.}, we prove 
$$
\MEM(p,s,n)\leq O(\LDC(p,1000s)\log n).
$$
In other words, if we have a decent $p$-probe LDC for encoding $O(s)$-bit strings, then we can use this to 
encode sets of size $s$ from a much larger universe $[n]$, at the expense of blowing up our 
data structure by only a factor of $\log n$. i
For instance, for $p=2$ probes we get $\MEM(2,s,n)\leq \exp(O(s))\log n$
from the Hadamard code, which is much better than the earlier $\exp(O(s\log n))$.
For $p=3$ probes, we get $\MEM(3,s,n)\leq \exp(\exp(\sqrt{\log s}))\log n$
from Efremenko's recent 3-probe LDC~\cite{efremenko:ldc}
(which improved Yekhanin's breakthrough construction~\cite{yekhanin:3ldc}).
Our construction relies heavily on the {\sc Membership} construction of~\cite{bmrv:bitvectorsj}.
Note that the near-tightness of the above upper and lower bounds implies that progress 
(meaning better upper and/or lower bounds) on locally decodable codes for any number of probes 
is \emph{equivalent} to progress on error-correcting data structures for $s$-out-of-$n$ {\sc Membership}.

\subsubsection{{\sc Inner product}}
In Section~\ref{secip} we analyze the inner product problem, where we are encoding $x\in\01^n$ 
and want to be able to compute the dot product $x\cdot y$ (mod 2), for any $y\in\01^n$ of weight at most $r$.
We first study the non-error-correcting setting, where we can prove nearly matching upper and
lower bounds (this is not the error-correcting setting, but provides something to compare it with).
Clearly, a trivial 1-probe data structure is to store the answers to all $B(n,r)$ possible queries separately.
In Section~\ref{ssecipnfault} we use a discrepancy argument from communication complexity
to prove a lower bound of about $B(n,r)^{1/p}$ on the length of $p$-probe data structures.
This shows that the trivial solution is essentially optimal if $p=1$.

We also construct various $p$-probe error-correcting data structures for inner product. 
For small $p$ and large $r$, their length is not much worse than the best non-error-correcting structures.
The upshot is that inner product is a problem where data structures can sometimes be
made error-correcting at little extra cost compared to the non-error-correcting case---admittedly, 
this is mostly because the non-error-correcting solutions for $\IP_{n,r}$ 
are already very expensive in terms of length.

\subsection{Related work}\label{ssecrelated}

Much work has of course been done on locally decodable codes, a.k.a.\ error-correcting data structures 
for the {\sc Membership} problem without constraints on the set size~\cite{trevisan:eccsurvey}. 
However, the error-correcting version of $s$-out-of-$n$ 
{\sc Membership} (``storing sparse tables'') or of other possible data structure problems 
has not been studied before.%
\footnote{Using the connection between information-theoretical private information
retrieval and locally decodable codes, one may derive some error-correcting
data structures from the PIR results of~\cite{cikrrw:private}. 
However, the resulting structures seem fairly weak.}
Here we briefly describe a number of other approaches to data structures in the presence of memory errors.
There is also much work on data structures with faulty \emph{processors}, but we will not discuss that here.

\paragraph{Fault-tolerant pointer-based data structures.}
Aumann and Bender~\cite{aumann&bender:ftdata} study fault-tolerant versions of \emph{pointer-based} 
data structures.  They define a pointer-based data structure as a directed graph where the edges are pointers,
and the nodes come in two types: information nodes carry real data, 
while auxiliary nodes carry auxiliary or structural data.
An \emph{error} is the destruction of a node and its outgoing edges.
They assume such an error is detected when accessing the node.
Even a few errors may be very harmful to pointer-based data structures: 
for instance, losing one pointer halfway a standard linked list means we lose the second half of the list.
They call a data structure \emph{$(d,g)$-fault-tolerant} (where $d$ is an integer that upper bounds the number of errors,
and $g$ is a function) if $f\leq d$ errors cause at most $g(f)$ information nodes to be lost.

Aumann and Bender present fault-tolerant \emph{stacks} with $g(f)=O(f)$, and fault-tolerant 
\emph{linked lists} and \emph{binary search trees} with $g(f)=O(f\log d)$, with only a constant-factor 
overhead in the size of the data structure, and small computational overhead.
Notice, however, that their error-correcting demands are much weaker than ours:
we require that \emph{no} part of the data is lost (every query should be answered 
with high success probability), even in the presence of a constant fraction of errors.
Of course, we pay for that in terms of the length of the data structure.

\paragraph{Faulty-memory RAM model.}
An alternative model of error-correcting data structures is the ``faulty-memory RAM model'',
introduced by Finocchi and Italiano~\cite{finocchi&italiano:faults}.
In this model, one assumes there are $O(1)$ incorruptible memory cells available.  
This is justified by the fact that CPU registers are much more robust than other kinds of memory. 
On the other hand, all other memory cells can be faulty---including the ones used by the algorithm
that is answering queries (something our model does not consider). 
The model assumes an upper bound $\Delta$ on the number of errors.

Finocchi, Grandoni, and Italiano described essentially optimal resilient algorithms for 
\emph{sorting} that work in $O(n\log n+\Delta^2)$ time with $\Delta$ up to about $\sqrt{n}$; 
and for \emph{searching} in $\Theta(\log n+\Delta)$ time.
There is a lot of recent work in this model:
J{\o}rgenson et al.~\cite{jmm:resilientpriority} study resilient \emph{priority queues},
Finocchi et al.~\cite{fgi:resilientsearch} study resilient \emph{search trees},
and Brodal et al.~\cite{bffgijmm:resilientdict} study resilient \emph{dictionaries}.
This interesting model allows for more efficient data structures than our model,
but its disadvantages are also clear: it assumes a small number of incorruptible cells, which may not be available
in many practical situations (for instance when the whole data structure is stored on a hard disk),
and the constructions mentioned above cannot deal well with a constant noise rate.

\section{The {\sc Membership} problem}\label{secmembership}

\subsection{Noiseless case: the BMRV data structure for {\sc Membership}}\label{secbmrvstructure}
Our error-correcting data structures for {\sc Membership} rely heavily on the construction 
of Buhrman et al.~\cite{bmrv:bitvectorsj}, whose relevant properties we sketch here. Their
structure is obtained using the probabilistic method.  Explicit but slightly less efficient
structures were subsequently given by Ta-Shma~\cite{tashma:storing}.
The BMRV-structure maps $x\in\01^n$ (of weight $\leq s$) to a string 
$y:=y(x)\in\01^{n'}$ of length $n'=\frac{100}{\eps^2}s\log n$ that can be decoded 
with one probe if $\delta=0$.  More precisely, for every $i\in[n]$ there is 
a set $P_i\subseteq[n']$ of size $|P_i|=\log(n)/\eps$ such that for every $x$ of weight $\leq s$:
\begin{equation}\label{eq1probemembership}
\Pr_{j\in P_i}[y_j=x_i]\geq 1-\eps,
\end{equation}
where the probability is taken over a uniform index $j\in P_i$.
For fixed $\eps$, the length $n'=O(s\log n)$ of the BMRV-structure is optimal up to a constant factor,
because clearly $\log{n\choose s}$ is a lower bound.

\subsection{Noisy case: 1 probe}
For the noiseless case, the BMRV data structure has information-theoretically optimal length $O(s\log n)$
and decodes with the minimal number of probes (one).
This can also be achieved in the error-correcting case if $s=1$:
then we just have the {\sc Equality} problem, for which see the remark following Definition~\ref{defftdatastr}. 
For larger $s$, one can observe that the BMRV-structure still works 
with high probability if $\delta\ll 1/s$: in that case the total number of errors is $\delta n'\ll \log n$,
so for each $i$, most bits in the $\Theta(\log n)$-set $P_i$ are uncorrupted.

\begin{theorem}[BMRV]
There exist $(1,\Omega(1/s),1/4)$-error-correcting data structures for {\sc Membership} of length $N=O(s\log n)$.
\end{theorem}

This only works if $\delta\ll 1/s$, which is actually close to optimal, as follows.
An $s$-bit LDC can be embedded in an error-correcting data structure for {\sc Membership},
hence it follows from Katz-Trevisan's~\cite[Theorem~3]{katz&trevisan:ldc} that there are no 
1-probe error-correcting data structures for {\sc Membership} if $s>1/(\delta(1-H(\eps)))$
(where $H(\cdot)$ denotes binary entropy).
In sum, there are 1-probe error-correcting data structures for {\sc Membership} of information-theoretically 
optimal length if $\delta\ll 1/s$.  In contrast, if $\delta\gg 1/s$ then there are no 1-probe error-correcting 
data structures at all, not even of exponential length.

\subsection{Noisy case: $p>1$ probes}

As we argued in the introduction, for fixed $\eps$ and $\delta$ there is an easy lower bound
on the length $N$ of $p$-probe error-correcting data structures for $s$-out-of-$n$ {\sc Membership}:
$$
N\geq\max\left(\LDC(p,s),\log\sum_{i=0}^s{n\choose i}\right).
$$
Our nearly matching upper bound, described below, uses the $\eps$-error 
data structure of~\cite{bmrv:bitvectorsj} for some small fixed $\eps$.
A simple way to obtain a $p$-probe error-correcting data structure is just to encode their 
$O(s\log n)$-bit string $y$ with the optimal $p$-probe LDC (with error $\eps'$, say),
which gives length $\LDC(p,O(s\log n))$. The one probe to $y$ is replaced by $p$ probes to the LDC.
By the union bound, the error probability of the overall construction is at most $\eps+\eps'$.
This, however, achieves more than we need: this structure enables us to recover $y_j$ for every $j$, 
whereas it would suffice if we were able to recover $y_j$ for most $j\in P_i$ (for each $i\in[n]$).

\paragraph{Definition of the data structure and decoder.}
To construct a shorter error-correcting data structure, we proceed as follows.
Let $\delta$ be a small constant (e.g.~$1/10000$);  
this is the noise level we want our final data structure for {\sc Membership} to protect against.
Consider the BMRV-structure for $s$-out-of-$n$ {\sc Membership}, with error probability at most 1/10.
Then $n'=10000s\log n$ is its length, and $b=10\log n$ is the size of each of the sets $P_i$.
Apply now a random permutation $\pi$ to $y$ (we show below that $\pi$ can be fixed to a specific permutation).
View the resulting $n'$-bit string as made up of $b=10\log n$ consecutive blocks of 
$1000s$ bits each. We encode each block with the optimal $(p,100\delta,1/100)$-LDC that encodes $1000s$ bits.
Let $\ell$ be the length of this LDC.  This gives overall length 
$$
N=10\ell\log n.
$$
The decoding procedure is as follows. Randomly choose a $k\in[b]$. This picks out one of the blocks. 
If this $k$th block contains exactly one $j\in P_i$ then recover $y_j$ from the (possibly corrupted) 
LDC for that block, using the $p$-probe LDC-decoder, and output $y_j$. 
If the $k$th block contains 0 or more than 1 elements from $P_i$, then output a uniformly random bit.

\paragraph{Analysis.}
Our goal below is to show that we can fix the permutation $\pi$ such that for at least $n/20$ of the indices $i\in[n]$, 
this procedure has good probability of correctly decoding $x_i$ (for all $x$ of weight $\leq s$).
The intuition is as follows. 
Thanks to the random permutation and the fact that $|P_i|$ equals the number of blocks, 
the expected intersection between $P_i$ and a block is exactly 1.
Hence for many $i\in[n]$, many blocks will contain exactly one index $j\in P_i$.
Moreover, for most blocks, their LDC-encoding won't have too many errors, hence we can recover $y_j$ using
the LDC-decoder for that block. Since $y_j=x_i$ for 90\%\ of the $j\in P_i$, we usually recover $x_i$.

To make this precise,
call $k\in[b]$ ``good for $i$'' if block $k$ contains \emph{exactly one} $j\in P_i$, 
and let $X_{ik}$ be the indicator random variable for this event.
Call $i\in[n]$ ``good'' if at least $b/4$ of the blocks are good for $i$ (i.e., $\sum_{k\in[b]} X_{ik}\geq b/4$), 
and let $X_i$ be the indicator random variable for this event.
The expected value (over uniformly random $\pi$) of each $X_{ik}$ is the probability that if we randomly 
place $b$ balls into $ab$ positions ($a$ is the block-size $1000s$), then there is exactly one ball 
among the $a$ positions of the first block, and the other $b-1$ balls are in the last $ab-a$ positions. 
This is
$$
\frac{a{ab-a\choose b-1}}{{ab\choose b}}=\frac{(ab-b)(ab-b-1)\cdots(ab-b-a+2)}{(ab-1)(ab-2)\cdots(ab-a+1)}
\geq\left(\frac{ab-b-a+2}{ab-a+1}\right)^{a-1}
\geq\left(1-\frac{1}{a-1}\right)^{a-1}.
$$
The righthand side goes to $1/e\approx 0.37$ with large $a$, so we can safely lower bound it by $3/10$.
Then, using linearity of expectation:
$$
\frac{3 b n}{10}\leq \Exp\left[\sum_{i\in[n],k\in[b]} X_{ik}\right]\leq b\cdot\Exp\left[\sum_{i\in[n]} X_i\right] + \frac{b}{4}\left(n-\Exp\left[\sum_{i\in[n]} X_i\right]\right),
$$
which implies 
$$
\Exp\left[\sum_{i\in[n]} X_i\right]\geq \frac{n}{20}.
$$
Hence we can fix one permutation $\pi$ such that at least $n/20$ of the indices $i$ are good.

For every index $i$, at least 90\%\ of all $j\in P_i$ satisfy $y_j=x_i$.
Hence for a good index $i$, with probability at least $1/4-1/10$ we will pick a $k$ such that the 
$k$th block is good for $i$ \emph{and} the unique $j\in P_i$ in the $k$th block satisfies $y_j=x_i$.
By Markov's inequality, the probability that the block that we picked has more than a
$100\delta$-fraction of errors, is less than $1/100$. If the fraction of errors is at most $100\delta$,
then our LDC-decoder recovers the relevant bit $y_j$ with probability $99/100$.
Hence the overall probability of outputting the correct value $x_i$ is at least
$$
\frac{3}{4}\cdot \frac{1}{2} + \left(\frac{1}{4}-\frac{1}{10}-\frac{1}{100}\right)\cdot \frac{99}{100}> \frac{51}{100}.
$$
We end up with an error-correcting data structure for {\sc Membership} for a universe of size $n/20$ instead of $n$
elements, but we can fix this by starting with the BMRV-structure for $20n$ bits.

We summarize this construction in a theorem:

\begin{theorem}\label{thldctoftmem}
If there exists a $(p,100\delta,1/100)$-LDC of length $\ell$ that encodes $1000s$ bits,
then there exists a $(p,\delta,49/100)$-error-correcting data structure of length $O(\ell\log n)$ 
for the $s$-out-of-$n$ {\sc Membership} problem.
\end{theorem}

The error and noise parameters of this new structure are not great, 
but they can be improved by more careful analysis.
We here sketch a better solution without giving all technical details.
Suppose we change the decoding procedure for $x_i$ as follows: pick $j\in P_i$
uniformly at random, decode $y_j$ from the LDC of the block where $y_j$ sits, and output the result.
There are three sources of error here: 
(1) the BMRV-structure makes a mistake (i.e., $j$ happens to be such that $y_j\neq x_i$),
(2) the LDC-decoder fails because there is too much noise on the LDC that we are decoding from,
(3) the LDC-decoder fails even though there is not too much noise on it.
The 2nd kind is hardest to analyze.  The adversary will do best if he puts just a bit more 
than the tolerable noise-level on the encodings of blocks that contain the most $j\in P_i$,
thereby ``destroying'' those encodings. 

For a random permutation, we expect that about $b/(e\cdot m!)$ of the $b$ blocks 
contain $m$ elements of $P_i$.  Hence about 1/65 of all blocks have 4 or more elements of $P_i$.
If the LDC is designed to protect against a $65\delta$-fraction of errors within one encoded block,
then with overall error-fraction $\delta$, the adversary has exactly enough noise to 
``destroy'' all blocks containing 4 or more elements of $P_i$.
The probability that our uniformly random $j$ sits in such a ``destroyed'' block is about 
$$
\sum_{m\geq 4}\frac{m}{b}\frac{b}{e\cdot m!}=\frac{1}{e}\left(\frac{1}{3!}+\frac{1}{4!}+\cdots\right)\approx 0.08.
$$
Hence if we set the error of the BMRV-structure to 1/10 and the error of the LDC to 1/100 (as above),
then the total error probability for decoding $x_i$ is less than 0.2 (of course we need to show that
we can fix a $\pi$ such that good decoding occurs for a good fraction of all $i\in[n]$).
Another parameter that may be adjusted is the block size, which we here took to be $1000s$.
Clearly, different tradeoffs between codelength, tolerable noise-level, and error probability are possible.

\section{The {\sc Inner product} problem}\label{secip}

\subsection{Noiseless case}\label{ssecipnfault}

Here we show bounds for {\sc Inner product}, first for the case where there is no noise ($\delta=0$).

\paragraph{Upper bound.}
Consider all strings $z$ of weight at most $\ceil{r/p}$.
The number of such $z$ is $B(n,\ceil{r/p})=\sum_{i=0}^{\ceil{r/p}}{n\choose i}\leq (e p n/r)^{r/p}$.
We define our codeword by writing down, for all $z$ in lexicographic order, the inner product $x\cdot z$ mod 2.
If we want to recover the inner product $x\cdot y$ for some $y$ of weight at most $r$, we write
$y=z_1+\cdots+z_p$ for $z_j$'s of weight at most $\ceil{r/p}$ and recover $x\cdot z_j$ for each $j\in[p]$,
using one probe for each.  Summing the results of the $p$ probes gives $x\cdot y$ (mod 2).
In particular, for $p=1$ probes, the length is $B(n,r)$.

\paragraph{Lower bound.}
To prove a nearly-matching lower bound, we use Miltersen's technique of relating 
a data structure to a two-party communication game~\cite{miltersen:unionsplitfind}.
We refer to~\cite{kushilevitz&nisan:cc} for a general introduction to communication complexity.
Suppose Alice gets string $x\in\01^n$, Bob gets string $y\in\01^n$ of weight $\leq r$, and they 
need to compute $x\cdot y$ (mod 2) with bounded error probability and minimal communication between them.
Call this communication problem $\IP_{n,r}$.
Let $B(n,r)=\sum_{i=0}^r {n\choose i}$ be the size of $Q$, i.e., the number of possible queries $y$.
The proof of our communication complexity lower bound below uses 
a fairly standard discrepancy argument, but we have not found this specific result
anywhere. For completeness we include a proof in Appendix~\ref{appdiscrip}.

\begin{theorem}\label{thipndbound}
Every communication protocol for $\IP_{n,r}$ with worst-case (or even average-case) success probability $\geq 1/2+\beta$ 
needs at least $\log(B(n,r)) - 2\log(1/2\beta)$ bits of communication.
\end{theorem}

Armed with this communication complexity bound we can lower bound data structure length:

\begin{theorem}\label{thipnrlb}
Every $(p,\eps)$-data structure for $\IP_{n,r}$ needs space 
$
\displaystyle N\geq \frac{1}{2}2^{(\log(B(n,r)) - 2\log(1/(1-2\eps)) - 1)/p}
$
\end{theorem}

\begin{proof}
We will use the data structure to obtain a communication protocol for $\IP_{n,r}$ that uses $p(\log(N)+1)+1$ 
bits of communication, and then invoke Theorem~\ref{thipndbound} to obtain the lower bound.

Alice holds $x$, and hence $\phi(x)$, while Bob simulates the decoder. Bob starts the communication.
He picks his first probe to the data structure and sends it over in $\log N$ bits. Alice sends back the 1-bit answer.
After $p$ rounds of communication, all $p$ probes have been simulated and Bob can give the same output as 
the decoder would have given. Bob's output will be the last bit of the communication. 
Theorem~\ref{thipndbound} now implies 
$
\displaystyle p(\log(N)+1)+1\geq \log(B(n,r)) - 2\log(1/(1-2\eps)).
$
Rearranging gives the bound on $N$.
\end{proof}

For fixed $\eps$, the lower bound is $N=\Omega\left(B(n,r)^{1/p}\right)$.
This is $\Omega((n/r)^{r/p})$, which (at least for small $p$)
is not too far from the upper bound of approximately $(e p n/r)^{r/p}$ mentioned above.
Note that in general our bound on $N$ is superpolynomial in $n$ whenever $p=o(r)$.
For instance, when $r=\alpha n$ for some constant $\alpha\in(0,1/2)$ then $N=\Omega(2^{nH(\alpha)/p})$,
which is non-trivial whenever $p=o(n)$. Finally, note that the proof technique also works if Alice's messages 
are longer than 1 bit (i.e., if the code is over a larger-than-binary alphabet).

\subsection{Noisy case}

\subsubsection{Constructions for {\sc Substring}}
One can easily construct error-correcting data structures for {\sc Substring}, which also suffice for {\sc Inner product}.
Note that since we are recovering $r$ bits, and each probe gives at most one bit of information,
by information theory we need at least about $r$ probes to the data structure.%
\footnote{$d/(\log(N)+1)$ probes in the case of \emph{quantum} decoders.}  
Our solutions below will use $O(r\log r)$ probes.
View $x$ as a concatenation $x=x^{(1)}\ldots x^{(r)}$ of $r$ strings of $n/r$ bits each 
(we ignore rounding for simplicity),
and define $\phi(x)$ as the concatenation of the Hadamard codes of these $r$ pieces.
Then $\phi(x)$ has length $N=r\cdot 2^{n/r}$.

If $\delta\geq 1/4r$ then the adversary could corrupt one of the $r$ Hadamard codes by 25\%\ noise, 
ensuring that some of the bits of $x$ are irrevocably lost even when we allow the full $N$ probes.
However, if $\delta\ll 1/r$ then we can recover each bit $x_i$ with small constant error probability by 2
probes in the Hadamard codeword where $i$ sits, and with error probability $\ll 1/r$ using $O(\log r)$ probes.
Hence we can compute $f(x,y)=x_y$ with error close to 0 using $p=O(r\log r)$ probes (or with $2r$ probes if $\delta\ll 1/r^2$).%
\footnote{It follows from Buhrman et al.~\cite{bnrw:robustqj} that if we allow a \emph{quantum} decoder, 
the factor of $\log r$ is not needed.}
This also implies that \emph{any} data structure problem where $f(x,q)$ depends on at most some fixed constant $r$ 
bits of $x$, has an error-correcting data structure of length $N=r\cdot 2^{n/r}$, $p=O(r\log r)$, and that works if $\delta\ll 1/r$.
Alternatively, we can take Efremenko's~\cite{efremenko:ldc} or Yekhanin's 3-probe LDC~\cite{yekhanin:3ldc}, 
and just decode each of the $r$ bits separately.
Using $O(\log r)$ probes to recover a bit with error probability
$\ll 1/r$, we recover the $r$-bit string $x_y$ using $p=O(r\log r)$ probes even if $\delta$ is a constant independent of~$r$.

\subsubsection{Constructions for {\sc Inner product}}
Going through the proof of~\cite{yekhanin:3ldc}, it is easy to see that it allows 
us to compute the parity of any set of $r$ bits from $x$ using at most $3r$ probes with error $\eps$, 
if the noise rate $\delta$ is at most $\eps/(3r)$ (just add the results of the 3 probes one 
would make for each bit in the parity).
To get error-correcting data structures even for small constant $p$ (independent of $r$), 
we can adapt the polynomial schemes from~\cite{bik:generalpir} to get the following theorem.
The details are given in Appendix~\ref{appftdataip}.

\begin{theorem}\label{thftdataip}
For every $p\geq 2$, there exists a $(p,\delta,p\delta)$-error-correcting data structure for
$\IP_{n,r}$ of length $N\leq p\cdot 2^{r (p-1)^2 n^{1/(p-1)}}$.
\end{theorem}

For the $p=2$ case, we get something simpler and better from the Hadamard code.
This code, of length $2^n$, actually allows us to compute $x\cdot y$ (mod 2)
for any $y\in\01^n$ of our choice, with 2 probes and error probability at most $2\delta$
(just probe $z$ and $y\oplus z$ for uniformly random $z\in\01^n$ and observe that 
$(x\cdot z)\oplus (x\cdot (y\oplus z))=x\cdot y$).
Note that for $r=\Theta(n)$ and $p=O(1)$, even non-error-correcting data structures
need length $2^{\Theta(n)}$ (Theorem~\ref{thipnrlb}).
This is an example where error-correcting data structures are not significantly
longer than the non-error-correcting kind.

\section{Future work}\label{secfuturework}

Many questions are opened up by our model of error-correcting data structures.  We mention a few:
\begin{itemize}
\item There are plenty of other natural data structure problems, such as {\sc Rank}, {\sc Predecessor},
versions of {\sc Nearest neighbor} etc.~\cite{miltersen:cellprobesurvey}.  
What about the length-vs-probes tradeoffs for their error-correcting versions?
The obvious approach is to put the best known LDC on top of the best known non-error-correcting data structures. 
This is not always optimal, though---for instance in the case of $s$-out-of-$n$ {\sc Membership} one can do
significantly better, as we showed.
\item It is often natural to assume that a memory cell contains not a bit, 
but some number from, say, a polynomial-size universe.  This is called the \emph{cell-probe} 
model~\cite{yao:tables}, in contrast to the \emph{bit-probe} model we considered here.
Probing a cell gives $O(\log n)$ bits at the same time, which can significantly improve the length-vs-probes tradeoff
and is worth studying.  Still, we view the bit-probe approach taken here as more fundamental than the cell-probe model.  
A $p$-probe cell-probe structure is a $O(p\log n)$-probe bit-probe structure, but not vice versa.  
Also, the way memory is addressed in actual computers in constant chunks of, say, 8 or 16 bits at a time, 
is closer in spirit to the bit-probe model than to the cell-probe model.
\item Zvi Lotker suggested to me the following connection with distributed computing.
Suppose the data structure is distributed over $N$ processors, each holding one bit.
Interpreted in this setting, an error-correcting data structure allows an honest party 
to answer queries about the encoded object while communicating with at most $p$ processors. 
The answer will be correct with probability $1-\eps$, even if up to a $\delta$-fraction of the $N$
processors are faulty or even malicious (the querier need not know where the faulty/malicious sites are).
\end{itemize}

\subsection*{Acknowledgments}
Thanks to Nitin Saxena for many useful discussions, 
to Harry Buhrman and Jaikumar Radhakrishnan for discussions about~\cite{bmrv:bitvectorsj},
to Zvi Lotker for the connection with distributed computation mentioned in Section~\ref{secfuturework},
to Peter Bro Miltersen for a pointer to~\cite{jmm:resilientpriority} and the faulty-memory RAM model,
and to Gabriel Moruz for sending me a copy of that paper.

\bibliographystyle{alpha}

\begin{thebibliography}{BNRW07}

\bibitem[AB96]{aumann&bender:ftdata}
Y.~Aumann and M.~Bender.
\newblock Fault-tolerant data structures.
\newblock In {\em Proceedings of 37th IEEE FOCS}, pages 580--589, 1996.

\bibitem[BFF{\etalchar{+}}07]{bffgijmm:resilientdict}
G.~Brodal, R.~Fagerberg, I.~Finocchi, F.~Grandoni, G.~Italiano,
  A.~J{\o}rgenson, G.~Moruz, and T.~M{\o}lhave.
\newblock Optimal resilient dynamic dictionaries.
\newblock In {\em Proceedings of 15th European Symposium on Algorithms (ESA)},
  pages 347--358, 2007.

\bibitem[BIK05]{bik:generalpir}
A.~Beimel, Y.~Ishai, and E.~Kushilevitz.
\newblock General constructions for information-theoretical {P}rivate
  {I}nformation {R}etrieval.
\newblock {\em Journal of Computer and System Sciences}, 72(2):247--281, 2005.

\bibitem[BMRV02]{bmrv:bitvectorsj}
H.~Buhrman, P.~B. Miltersen, J.~Radhakrishnan, and S.~Venkatesh.
\newblock Are bitvectors optimal?
\newblock {\em SIAM Journal on Computing}, 31(6):1723--1744, 2002.
\newblock Earlier version in STOC'00.

\bibitem[BNRW07]{bnrw:robustqj}
H.~Buhrman, I.~Newman, H.~R{\"o}hrig, and R.~{de} Wolf.
\newblock Robust polynomials and quantum algorithms.
\newblock {\em Theory of Computing Systems}, 40(4):379--395, 2007.

\bibitem[CIK{\etalchar{+}}01]{cikrrw:private}
R.~Canetti, Y.~Ishai, R.~Kumar, M.~Reiter, R.~Rubinfeld, and R.~Wright.
\newblock Selective private function evaluation with applications to private
  statistics.
\newblock In {\em Proceedings of 20th Annual ACM Symposium on Principles of
  Distributed Computing (PODC)}, pages 293--304, 2001.

\bibitem[Efr08]{efremenko:ldc}
K.~Efremenko.
\newblock 3-query locally decodable codes of subexponential length.
\newblock Technical report, ECCC Report TR08--069, 2008.

\bibitem[FGI07]{fgi:resilientsearch}
I.~Finocchi, F.~Grandoni, and G.~Italiano.
\newblock Resilient search trees.
\newblock In {\em Proceedings of 18th ACM-SIAM SODA}, pages 547--553, 2007.

\bibitem[FI04]{finocchi&italiano:faults}
I.~Finocchi and G.~Italiano.
\newblock Sorting and searching in the presence of memory faults (without
  redundancy).
\newblock In {\em Proceedings of 36th ACM STOC}, pages 101--110, 2004.

\bibitem[FKS84]{fks:sparsetable}
M.~Fredman, M.~Koml{\'o}s, and E.~Szemer{\'e}di.
\newblock Storing a sparse table with {$O(1)$} worst case access time.
\newblock {\em Journal of the ACM}, 31(3):538--544, 1984.

\bibitem[JMM07]{jmm:resilientpriority}
A.~G. J{\o}rgenson, G.~Moruz, and T.~M{\o}lhave.
\newblock Resilient priority queues.
\newblock In {\em Proceedings of 10th International Workshop on Algorithms and
  Data Structures (WADS)}, volume 4619 of {\em Lecture Notes in Computer
  Science}, 2007.

\bibitem[KN97]{kushilevitz&nisan:cc}
E.~Kushilevitz and N.~Nisan.
\newblock {\em Communication Complexity}.
\newblock Cambridge University Press, 1997.

\bibitem[KT00]{katz&trevisan:ldc}
J.~Katz and L.~Trevisan.
\newblock On the efficiency of local decoding procedures for error-correcting
  codes.
\newblock In {\em Proceedings of 32nd ACM STOC}, pages 80--86, 2000.

\bibitem[KW04]{kerenidis&wolf:qldcj}
I.~Kerenidis and R.~{de} Wolf.
\newblock Exponential lower bound for 2-query locally decodable codes via a
  quantum argument.
\newblock {\em Journal of Computer and System Sciences}, 69(3):395--420, 2004.

\bibitem[Mil94]{miltersen:unionsplitfind}
P.~B. Miltersen.
\newblock Lower bounds for {U}nion-{S}plit-{F}ind related problems on random
  access machines.
\newblock In {\em Proceedings of 26th ACM STOC}, pages 625--634, 1994.

\bibitem[Mil99]{miltersen:cellprobesurvey}
P.~B. Miltersen.
\newblock Cell probe complexity - a survey. {I}nvited paper at \emph{Advances
  in Data Structures} workshop.
\newblock Available at Miltersen's homepage, 1999.

\bibitem[MS77]{MacWilliamsS77}
F.~MacWilliams and N.~Sloane.
\newblock {\em The Theory of Error-Correcting Codes}.
\newblock North-Holland, 1977.

\bibitem[RSV02]{rsv:qsetmembershipj}
J.~Radhakrishnan, P.~Sen, and S.~Venkatesh.
\newblock The quantum complexity of set membership.
\newblock {\em Algorithmica}, 34(4):462--479, 2002.
\newblock Earlier version in FOCS'00.

\bibitem[Tre04]{trevisan:eccsurvey}
L.~Trevisan.
\newblock Some applications of coding theory in computational complexity.
\newblock {\em Quaderni di Matematica}, 13:347--424, 2004.

\bibitem[TS02]{tashma:storing}
A.~Ta-Shma.
\newblock Storing information with extractors.
\newblock {\em Information Processing Letters}, 83(5):267--274, 2002.

\bibitem[vL98]{lint:ecc}
J.~H. van Lint.
\newblock {\em Introduction to Coding Theory}.
\newblock Springer, third edition, 1998.

\bibitem[Yao77]{yao:unified}
A.~C-C. Yao.
\newblock Probabilistic computations: Toward a unified measure of complexity.
\newblock In {\em Proceedings of 18th IEEE FOCS}, pages 222--227, 1977.

\bibitem[Yao81]{yao:tables}
A.~C-C. Yao.
\newblock Should tables be sorted?
\newblock {\em Journal of the ACM}, 28(3):615--628, 1981.

\bibitem[Yek07]{yekhanin:3ldc}
S.~Yekhanin.
\newblock Towards 3-query locally decodable codes of subexponential length.
\newblock In {\em Proceedings of 39th ACM STOC}, pages 266--274, 2007.

\end{thebibliography}

\newcommand{\etalchar}[1]{$^{#1}$}

\appendix

\section{Proof of Theorem~\ref{thipndbound}}\label{appdiscrip}

Let $\mu$ be the uniform input distribution: 
each $x$ has probability $1/2^n$ and each $y$ of weight $\leq r$ has probability $1/B(n,r)$.
We show a lower bound on the communication $c$ of \emph{deterministic} protocols that compute $\IP_{n,r}$ 
with $\mu$-probability at least $1/2+\beta$.  By Yao's principle~\cite{yao:unified}, 
this lower bound then also applies to randomized protocols.

Consider a deterministic $c$-bit protocol. Assume the last bit communicated is the output bit.
It is well-known that this partitions the input space into \emph{rectangles} $R_1,\ldots,R_{2^c}$, 
where $R_i=A_i\times B_i$, and the protocol gives the same output bit $a_i$ for each 
$(x,y)\in R_i$.\footnote{\cite[Section~1.2]{kushilevitz&nisan:cc}.
The number of rectangles may be smaller than $2^c$, but we can always add empty ones.}
The \emph{discrepancy} of rectangle $R=A\times B$ under $\mu$ is
the difference between the weight of the 0s and the 1s in that rectangle:
$$
\delta_\mu(R)=\left|\mu(R\cap\IP_{n,r}^{-1}(1))-\mu(R\cap\IP_{n,r}^{-1}(0))\right|
$$
We can show for every rectangle that its discrepancy is not very large:

\begin{lemma}
$\displaystyle\delta_\mu(R)\leq\frac{\sqrt{|R|}}{\sqrt{2^n}B(n,r)}$.
\end{lemma}

\begin{proof}
Let $M$ be the $2^n\times B(n,r)$ matrix whose $(x,y)$-entry is $(-1)^{{\rm IP}_{n,r}(x,y)}=(-1)^{x\cdot y}$.
It is easy to see that $M^T M=2^n I$, where $I$ is the $B(n,r)\times B(n,r)$ identity matrix.  
This implies, for any $v\in\mathbb{R}^{B(n,r)}$
$$
\norm{M v}^2=(Mv)^T\cdot(Mv)=v^TM^TM v=2^n v^T v=2^n\norm{v}^2.
$$
Let $R=A\times B$, $v_A\in\01^{2^n}$ and $v_B\in\01^{B(n,r)}$ be the
characteristic (column) vectors of the sets $A$ and $B$.
Note that $\norm{v_A}=\sqrt{|A|}$ and $\norm{v_B}=\sqrt{|B|}$.
The sum of $M$-entries in $R$ is $\sum_{a\in A, b\in B}M_{ab}=v_A^T M v_B$.
We can bound this using Cauchy-Schwarz:
$$
|v_A^T M v_B|\leq \norm{v_A}\cdot \norm{M v_B}=\norm{v_A}\cdot \sqrt{2^n}\norm{v_B}=\sqrt{|A|\cdot|B|\cdot 2^n}.
$$
Observing that $\delta_\mu(R)=|v_A^T M v_B|/(2^n B(n,r))$ and $|R|=|A|\cdot|B|$ concludes the proof.
\end{proof}

Define the success and failure probabilities (under $\mu$) of the protocol as
$$
P_s=\sum_{i=1}^{2^c}\mu(R_i\cap\IP_{n,r}^{-1}(a_i))\mbox{ \ and \ }P_f=\sum_{i=1}^{2^c}\mu(R_i\cap\IP_{n,r}^{-1}(1-a_i))
$$
Then
\begin{eqnarray*}
2\beta & \leq & P_s-P_f\\
      &   =  & \sum_i\mu(R_i\cap\IP_{n,r}^{-1}(a_i))-\mu(R_i\cap\IP_{n,r}^{-1}(1-a_i))\\
      & \leq & \sum_i\left|\mu(R_i\cap\IP_{n,r}^{-1}(a_i))-\mu(R_i\cap\IP_{n,r}^{-1}(1-a_i))\right|\\
      &   =  & \sum_i\delta_\mu(R_i)
       ~\leq~  \frac{\sum_i\sqrt{|R_i|}}{\sqrt{2^n}B(n,r)}
       ~\leq~  \frac{\sqrt{2^c}\sqrt{\sum_i|R_i|}}{\sqrt{2^n}B(n,r)}
         ~=~   \sqrt{2^c/B(n,r)},
\end{eqnarray*}
where the last inequality is Cauchy-Schwarz and the last equality holds because $\sum_i|R_i|$ 
is the total number of inputs, which is $2^n B(n,r)$.

Rearranging gives $2^c\geq (2\beta)^2 B(n,r)$, hence $c\geq \log(B(n,r)) - 2\log(1/2\beta)$.

\section{Proof of Theorem~\ref{thftdataip}}\label{appftdataip}

Here we construct $p$-probe error-correcting data structures for the inner product problem,
inspired by the approach to locally decodable codes of~\cite{bik:generalpir}.
Let $d$ be an integer to be determined later.
Pick $m=\ceil{d n^{1/d}}$.  Then ${m\choose d}\geq n$, so there exist $n$ distinct sets
$S_1,\ldots,S_n\subseteq[m]$, each of size $d$.
For each $x\in\01^n$, define an $m$-variate polynomial $p_x$ of degree $d$ over $\mathbb{F}_2$ by
$$
p_x(z_1,\ldots,z_m)=\sum_{i=1}^n x_i\prod_{j\in S_i}z_j.
$$
Note that if we identify $S_i$ with its $m$-bit characteristic vector, then $p_x(S_i)=x_i$.
For $z^{(1)},\ldots,z^{(r)}\in\01^m$, define an $rm$-variate polynomial $p_{x,r}$ over $\mathbb{F}_2$ by
$$
p_{x,r}(z^{(1)},\ldots,z^{(r)})=\sum_{j=1}^r p_x(z^{(j)}).
$$
This polynomial $p_{x,r}(z)$ has $r m$ variables, degree $d$, and allows us to evaluate parities 
of any set of $r$ of the variables of $x$: if $y\in\01^n$ (of weight $r$) has its 1-bits at
positions $i_1,\ldots,i_r$, then 
$$
p_{x,r}(S_{i_1},\ldots,S_{i_r})=\sum_{j=1}^r x_{i_j}=x\cdot y \mbox{\ (mod 2)}.
$$
To construct an error-correcting data structure for $\IP_{n,r}$,
it thus suffices to give a structure that enables us to evaluate $p_{x,r}$ at any point $w$ of our choice.%
\footnote{If we also want to be able to compute $x\cdot y$ (mod 2) for $|y|<r$, 
we can just add a dummy 0 as $(n+1)$st variable to $x$, and use its index $r-|y|$ times as inputs to $p_{x,r}$.}

Let $w\in\01^{r m}$.  Suppose we ``secret-share'' this into $p$ pieces $w^{(1)},\ldots,w^{(p)}\in\01^{r m}$
which are uniformly random subject to the constraint $w=w^{(1)}+\cdots+w^{(p)}$.
Now consider the $prm$-variate polynomial $q_{x,r}$ defined by
\begin{equation}\label{eqqxr1}
q_{x,r}(w^{(1)},\ldots,w^{(p)})=p_{x,r}(w^{(1)}+\cdots+w^{(p)}).
\end{equation}
Each monomial $M$ in this polynomial has at most~$d$ variables. If we pick $d=p-1$, then
for every $M$ there will be a $j\in[p]$ such that $M$ does not contain variables from $w^{(j)}$.
Assign all such monomials to a new polynomial 
$q^{(j)}_{x,r}$, which is independent of $w^{(j)}$.
This allows us to write
\begin{equation}\label{eqqxr2}
q_{x,r}(w^{(1)},\ldots,w^{(p)})=q^{(1)}_{x,r}(w^{(2)},\ldots,w^{(p)})+\cdots+q^{(p)}_{x,r}(w^{(1)},\ldots,w^{(p-1)}).
\end{equation}
Note that each $q^{(j)}_{x,r}$ has domain of size $2^{(p-1)rm}$.
The data structure is defined as the concatenation, 
for all $j\in[p]$, of the values of $q^{(j)}_{x,r}$ on all possible inputs.
This has length 
$$
N=p\cdot 2^{(p-1) r m}=p\cdot 2^{r (p-1)^2 n^{1/(p-1)}}.
$$  
This length is $2^{O(rn^{1/(p-1)})}$ for $p=O(1)$.

Answering a query works as follows: the decoder would like to evaluate $p_{x,r}$ on some point $w\in\01^{r m}$.
He picks $w^{(1)},\ldots,w^{(p)}$ as above, and for all $j\in[p]$,  
in the $j$th block of the code probes the point $z^{(1)},\ldots,z^{(j-1)},z^{(j+1)},\ldots,z^{(p)}$. 
This, if uncorrupted, returns the value of $q^{(j)}_{x,r}$ at that point.
The decoder outputs the sum of his $p$ probes (mod 2). 
If none of the probed bits were corrupted, 
then the output is $p_{x,r}(w)$ by Eqs.~(\ref{eqqxr1}) and~(\ref{eqqxr2}).
Note that the probe within the $j$th block is uniformly random in that block, so its error probability 
is exactly the fraction $\delta_j$ of errors in the $j$th block.  
Hence by the union bound, the total error probability is at most $\sum_{j=1}^p\delta_j$.
If the overall fraction of errors in the data structure is at most $\delta$, then we have 
$\frac{1}{p}\sum_{j=1}^p\delta_j\leq\delta$, hence the total error probability is at most $p\delta$.

\end{document}